%% file: bam954_FH_CL.tex
\documentclass[a4paper,11pt]{article}
\usepackage{jheppub}
\usepackage[abs]{overpic}
\usepackage{warpcol}
\usepackage[T1]{fontenc}

\usepackage{lineno}
\usepackage{epstopdf}
\usepackage[normalem]{ulem}

\newcommand{\EE}{e^+e^-}

\usepackage{rotating, booktabs}

\arxivnumber{1234.56789}

\title{\boldmath Measurement of the branching fractions of $\chi_{cJ} \to \pi^{+}\pi^{-}\pi^{0}\pi^{0}$ via $\psi(3686) \to \gamma\chi_{cJ}$}

\collaborationImg{\includegraphics[width=.12\textwidth,origin=c,angle=90]{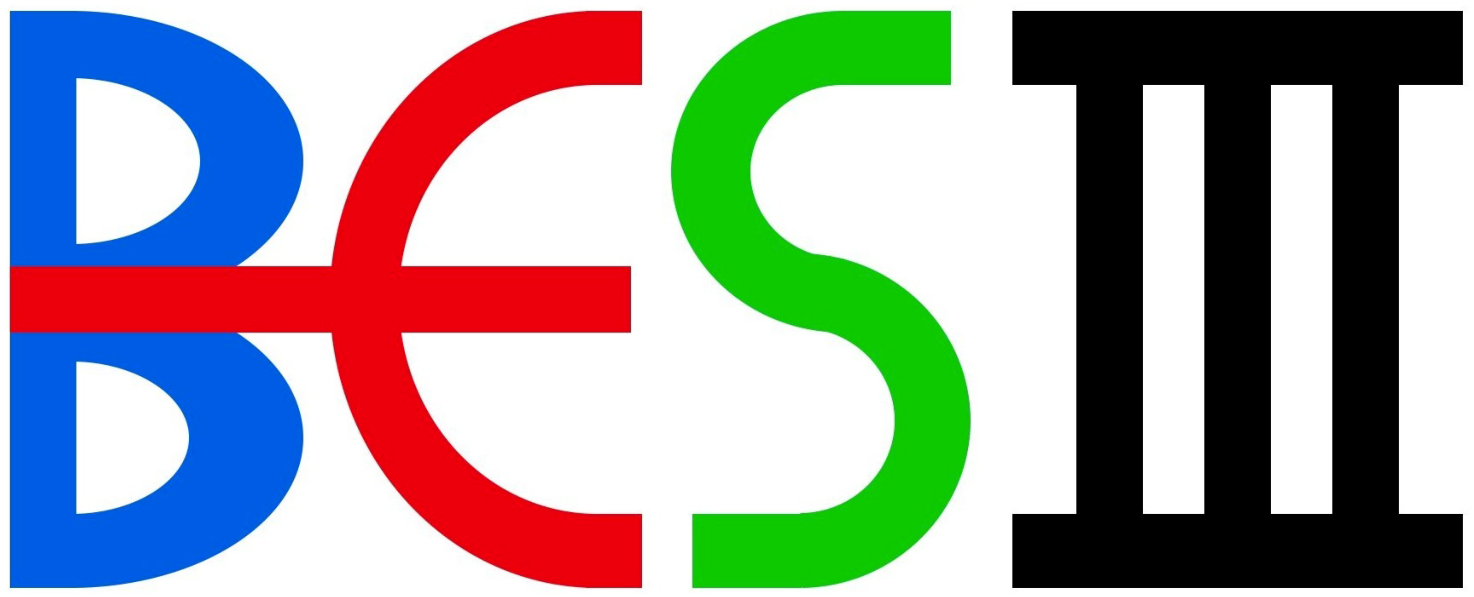}}

\collaboration{BESIII Collaboration}


\emailAdd{besiii-publications@ihep.ac.cn}

\abstract{
Using $(2712.4\pm14.3)\times 10^6$ $\psi(3686)$ events collected with
the BESIII detector operating at BEPCII, the branching fractions
of $\chi_{cJ}\to\pi^+\pi^-\pi^0\pi^0$ ($J=0,~1,~2$) are measured via
the radiative transition $\psi(3686)\to\gamma\chi_{cJ}$. The results
are $\mathcal{B}(\chi_{c0} \to \pi^{+}\pi^{-}\pi^{0}\pi^{0}) = (3.10
\pm 0.01 \pm 0.14) \times 10^{-2}$, $\mathcal{B}(\chi_{c1} \to
\pi^{+}\pi^{-}\pi^{0}\pi^{0}) = (1.16 \pm 0.01 \pm 0.05) \times
10^{-2}$, and $\mathcal{B}(\chi_{c2} \to \pi^{+}\pi^{-}\pi^{0}\pi^{0})
= (1.92 \pm 0.01 \pm 0.08) \times 10^{-2}$, where the first
uncertainties are statistical and the second systematic.  The dominant
intermediate states are found to be $\chi_{cJ}\to\rho^+\rho^-$.  These
results supersede the previous most precise measurements and provide
significantly improved precision.}

\begin{document}
\newcommand{\BESIIIorcid}[1]{\href{https://orcid.org/#1}{\hspace*{0.1em}\raisebox{-0.45ex}{\includegraphics[width=1em]{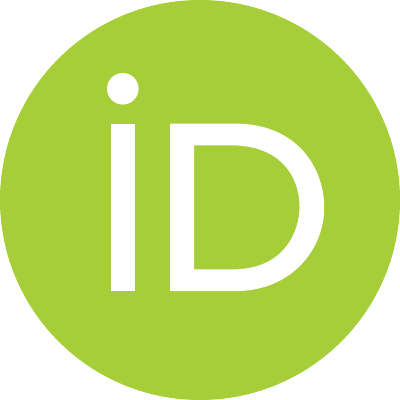}}}} 
\maketitle
\flushbottom


\section{Introduction}
\label{sec:intro}

Charmonium decays have attracted widespread interest as an excellent
laboratory for studying quark-gluon dynamics at relatively low
energy. Precise measurements of the decay properties of
conventional charmonium states, particularly the P-wave triplet states
$\chi_{cJ} (J=0,1,2)$, remain an important research within the BESIII
experiment and in the broader field of charmonium physics
\cite{bes3review2022}. Many decay modes remain unobserved. Even for
the known decay modes, some measured branching fractions still need to
be improved due to limited statistics in earlier
measurements~\cite{2008CLEO}. Previous theoretical studies indicate
that the color-octet-mechanism plays a role in the decay of the
$P$-wave charmonium states \cite{1997HWHuang, 1996APetrelli,
1997JBolz, 2000SMHWong1}.  In order to build a comprehensive
understanding of the $P$-wave dynamics, more precise experimental
measurements for $\chi_{cJ}$ are necessary.  Although $\chi_{cJ}$
states are hard to produce in $e^{+}e^{-}$ collisions via the
annihilation process, they can be abundantly produced in the radiative transitions $\psi(3686) \to \gamma\chi_{cJ}$ with a sizable
branching fraction ($\sim$9\,\%)~\cite{PDG}. Currently, the world’s
largest $\psi(3686)$ data sample~\cite{allpsi} produced directly in
$\EE$ collisions has been collected at BESIII. This provides an
excellent opportunity to further investigate the decay dynamics of
$\chi_{cJ}$.

In this work, we measure the branching fractions of $\chi_{cJ} \to \pi^{+}\pi^{-}\pi^{0}\pi^{0}$ via the radiative transition processes of $\psi(3686)\to\gamma\chi_{cJ}$. We also investigate the intermediate states contributing to these decays.

\section{BESIII detector and Monte Carlo simulation}

The BESIII detector is a magnetic spectrometer~\cite{besiii} located
at the Beijing Electron Positron Collider (BEPCII)~\cite{bepcii}. The
cylindrical core of the BESIII detector consists of a helium-based
multilayer drift chamber (MDC)~\cite{MDC}, a plastic scintillator
time-of-flight system (TOF), and a CsI (Tl) electromagnetic
calorimeter (EMC), which are all enclosed in a superconducting
solenoidal magnet, providing a 1.0~T magnetic field. The solenoid is
supported by an octagonal flux-return yoke with resistive plate
chamber muon identifier modules interleaved with steel. The acceptance
of charged particles and photons is 93\,\% over the $4\pi$ solid
angle. The charged-particle momentum resolution at $1~{\rm GeV}/c$ is
$0.5\,\%$, and the $\mathrm{d}E/\mathrm{d}x$ resolution is $6\,\%$ for
the electrons from Bhabha scattering. The EMC measures photon energies
with a resolution of $2.5\,\%$ ($5\,\%$) at $1$~GeV in the barrel (end
cap) region. The time resolution of the TOF barrel section is 68~ps,
while that of the end cap section was 110~ps. The end cap TOF system
was upgraded in 2015 with multi-gap resistive plate chamber
technology, providing a time resolution of 60~ps~\cite{etof1, etof2,
etof3}, which is relevant for 83\,$\%$ of the data used in this
analysis.

This analysis is based on the $\psi(3686)$ data sample collected with
the BESIII detector. The data were taken at the $\psi(3686)$ resonance
peak during three running periods in 2009, 2012, and 2021, with
integrated luminosities of $161.63~\text{pb}^{-1}$,
$506.92~\text{pb}^{-1}$, and $3208.5~\text{pb}^{-1}$,
respectively. The number of $\psi(3686)$ events is determined to
be $(2.712\ \pm\ 0.014) \times 10^9$ via an analysis of inclusive
hadronic decays \cite{allpsi}. The data samples were reconstructed
with BOSS~\cite{boos}.

The BESIII detector is simulated using the GEANT4-based~\cite{geant4}
simulation software BOOST~\cite{Boost}, which includes the geometric
and material description of the BESIII detector, the detector response
and digitization models, as well as a record of the detector running
conditions and performances.

To determine the detection efficiencies, exclusive Monte Carlo samples
are generated. The production process $\psi(3686) \to \gamma\chi_{cJ}$
follows a pure electric dipole ($\mathrm{E1}$)
transition~\cite{gtheta01}, while the subsequent decays $\chi_{cJ} \to
\pi^{+}\pi^{-}\pi^{0}\pi^{0}$ are generated with a uniform phase-space
(PHSP) distribution. To study the intermediate states, dedicated MC
samples are generated using the VSS model~\cite{BesEvt}, which
provides an accurate simulation of $\rho \to \pi\pi$ decay dynamics.
For the inclusive $\psi(3686)$ sample, the $\psi(3686)$ resonance is
simulated using KKMC~\cite{KKMC1,KKMC2}, with known decays handled by
BesEvtGen~\cite{BesEvt} using branching fractions from the Particle
Data Group (PDG)~\cite{PDG}, and unknown decays are modeled by
LUNDCHARM~\cite{LUNDCHARM}. These inclusive MC samples, comprising 2.7
billion $\psi(3686)$ events~\cite{allpsi}, are used to analyze
background contributions and to optimize the event selection criteria
in order to enhance signal efficiency and reduce background
contamination.

\section{Event selection and background analysis}
\subsection{General event selection}

Charged tracks detected in the MDC are required to be within
$|\cos\theta| < 0.93$, where the polar angle $\theta$ is defined with
respect to the $z$-axis, the symmetry axis of the MDC. The distance
of the closest approach to the interaction point, $|V_z|$ , must be less
than 10 cm in the $z$-direction and the distance $|V_{xy}|$ must be
less than 1 cm in the plane perpendicular to $z$, for all charged
tracks.

Particle identification (PID) for charged tracks combines the
$\mathrm{d}E/\mathrm{d}x$ measurements in the MDC and the information for TOF to form
likelihoods $\mathcal{L}(h)$ for each hadron
hypothesis $h$ ($h = p, K, \pi$). Tracks are identified as pions if the pion
hypothesis has the greatest likelihood ($\mathcal{L}(\pi) >
\mathcal{L}(K)$ and $\mathcal{L}(\pi) > \mathcal{L}(p)$).

Photon candidates are identified using showers in the EMC. The
deposited energy of each shower must exceed $25~\text{MeV}$ in the barrel region ($|\cos\theta| < 0.80$) and  $50~\text{MeV}$ in the end cap region ($0.86 < |\cos\theta| <
0.92$). To exclude showers that originate from charged tracks, the
angle between the position of each shower in the EMC and the closest
extrapolated charged track must be greater than $10^\circ$. To
suppress electronic noise and showers unrelated to the event, the
difference between the EMC time and the event start time is required
to be within $[0, 700]~\text{ns}$.

Since the final state under study is
$\gamma\pi^{+}\pi^{-}\pi^{0}\pi^{0}$ (with each $\pi^{0}$
reconstructed from two photons), candidate events are required to contain two charged tracks, identified as pions with zero net charge and at least five photons.

\subsection{Further event selection and background study}

In order to reduce background and improve the mass resolution, a six
constraint (6C) kinematic fit is performed, constraining the final
state energy-momentum to the total initial four-momentum of the
colliding beams and the masses of the two $\pi^0$ candidates to the
known $\pi^0$ mass and providing a goodness-of-fit, $\chi^{2}_{\text{6C}}$, is then obtained. If there is more than one $\pi^0\pi^0$
combinations, the one with the smallest $\chi^{2}_{\text{6C}}$ is
selected.  Furthermore, a requirement of $\chi^{2}_{\text{6C}} < 60$
is imposed. This criterion was determined by optimizing the
Figure-of-Merit (FOM), defined as $\text{FOM} =
S_{\text{MC}}/\sqrt{(S_{\text{MC}}+B_{\text{incMC}})}$. Here,
$S_{\text{MC}}$ denotes the normalized event yield from the signal MC
sample, and $B_{\text{incMC}}$ represents the estimated background
yield derived from the inclusive MC sample. The sum
$(S_{\text{MC}}+B_{\text{incMC}})$ corresponds to the total normalized
yield from the inclusive MC sample.  To suppress background from
events with more or less than five photons, four-constraint
energy-momentum conservation kinematic fits are performed for
$6\gamma\pi^{+}\pi^{-}$ (if available), $5\gamma\pi^{+}\pi^{-}$, and
$4\gamma\pi^{+}\pi^{-}$ hypotheses corresponding to
$\chi^{2}_{\text{more}}$, $\chi^{2}_{\text{norm}}$, and
$\chi^{2}_{\text{less}}$, respectively, and $\chi^{2}_{\text{norm}} <
\chi^{2}_{\text{more}}$ and $\chi^{2}_{\text{norm}} <
\chi^{2}_{\text{less}}$ are required.

After these selections, potential backgrounds are investigated using
the inclusive MC sample. The main backgrounds are found to originate
from $\psi(3686) \to \pi^{0}\pi^{0}J/\psi$, $\pi^{+}\pi^{-}J/\psi$,
$\pi^+\pi^-\pi^0 J/\psi$ and the peaking background $\chi_{cJ}
\to K^0_{\mathrm{S}}K^0_{\mathrm{S}}$ with one $K^0_{\mathrm{S}}$
decaying to $\pi^+\pi^-$ and the other to $\pi^0\pi^0$.  The $\pi\pi
J/\psi$ backgrounds are vetoed by requiring the $\pi\pi$ recoil mass
$M_{\text{R}}$ to be outside the $J/\psi$ signal region:
$M_{\text{R}}(\pi^0\pi^0) > 3.07$ GeV/$c^2$ and
3.09~GeV/$c^2$~$<~M_{\text{R}}(\pi^+\pi^-) <$ 3.10 GeV/$c^2$. Events
with $ \pi^+\pi^-\pi^0$ (predominantly from $\eta J/\psi$), are vetoed
by requiring the $\pi^+\pi^-\pi^0$ invariant mass
$M(\pi^+\pi^-\pi^0_{\text{high}})$ to lie outside the range 3.09 to
3.18 GeV/$c^2$, where $\pi^0_{\text{high}}$ refers to the
higher-momentum $\pi^0$ candidate among the two reconstructed.
Additionally, the peaking background from
$\chi_{cJ}\,\to\,K^0_{\mathrm{S}} K^0_{\mathrm{S}}$ is vetoed by
excluding events within the 2-dimensional mass window: $M(\pi^+\pi^-)
\in (0.445, 0.515)$ GeV/$c^2$ and $M(\pi^0\pi^0) \in (0.475, 0.520)$
GeV/$c^2$. All veto windows and mass requirements use widths of
$\pm3\sigma$, where the $\sigma$s are determined by double-Gaussian
fits to the corresponding mass peaks in the exclusive MC sample.

\begin{figure}[htbp]
\centering
\includegraphics[width=0.77\textwidth]{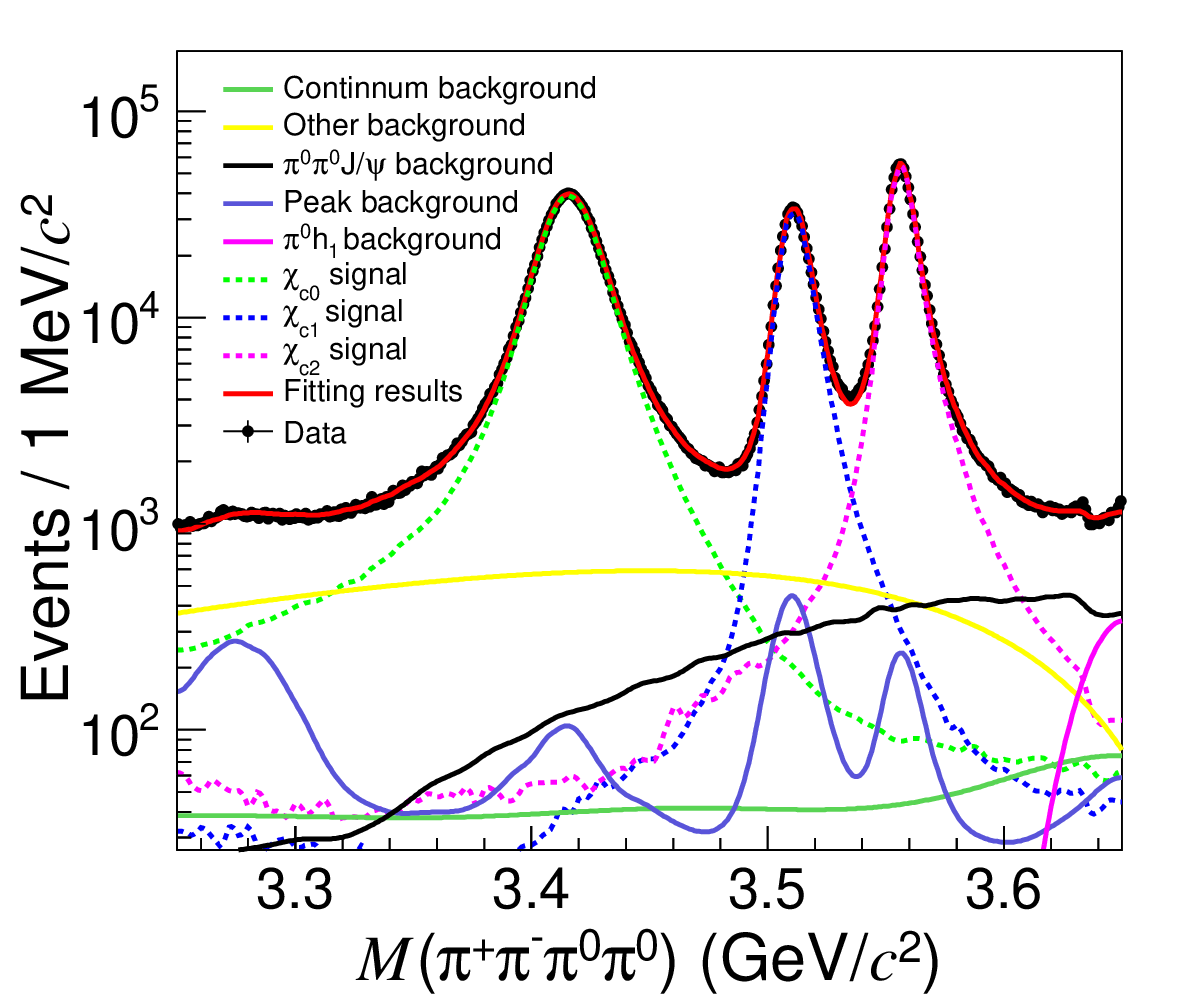}
\caption{Distribution of $M(\pi^{+}\pi^{-}\pi^{0}\pi^{0})$ in data and
the background contributions from the MC. The data points with error bars
are shown in black. The total fit result is depicted by the red solid
curve. The different $\chi_{cJ}$ states ($J=0,1,2$) are indicated by
dashed curves in distinct colors. The individual background components
are: the pink solid curve, dominated by backgrounds with $\pi^0h_1$;
the blue solid curve, accounting for the peaking backgrounds; the
black solid curve, primarily from $\pi^0\pi^0 J/\psi$ events; the
yellow solid curve, representing a smooth description of other
backgrounds; and the green solid curve, for the continuum background.}
\label{FIGDBA}
\end{figure}

\begin{figure}[htbp]
\centering
\includegraphics[width=0.32\textwidth]{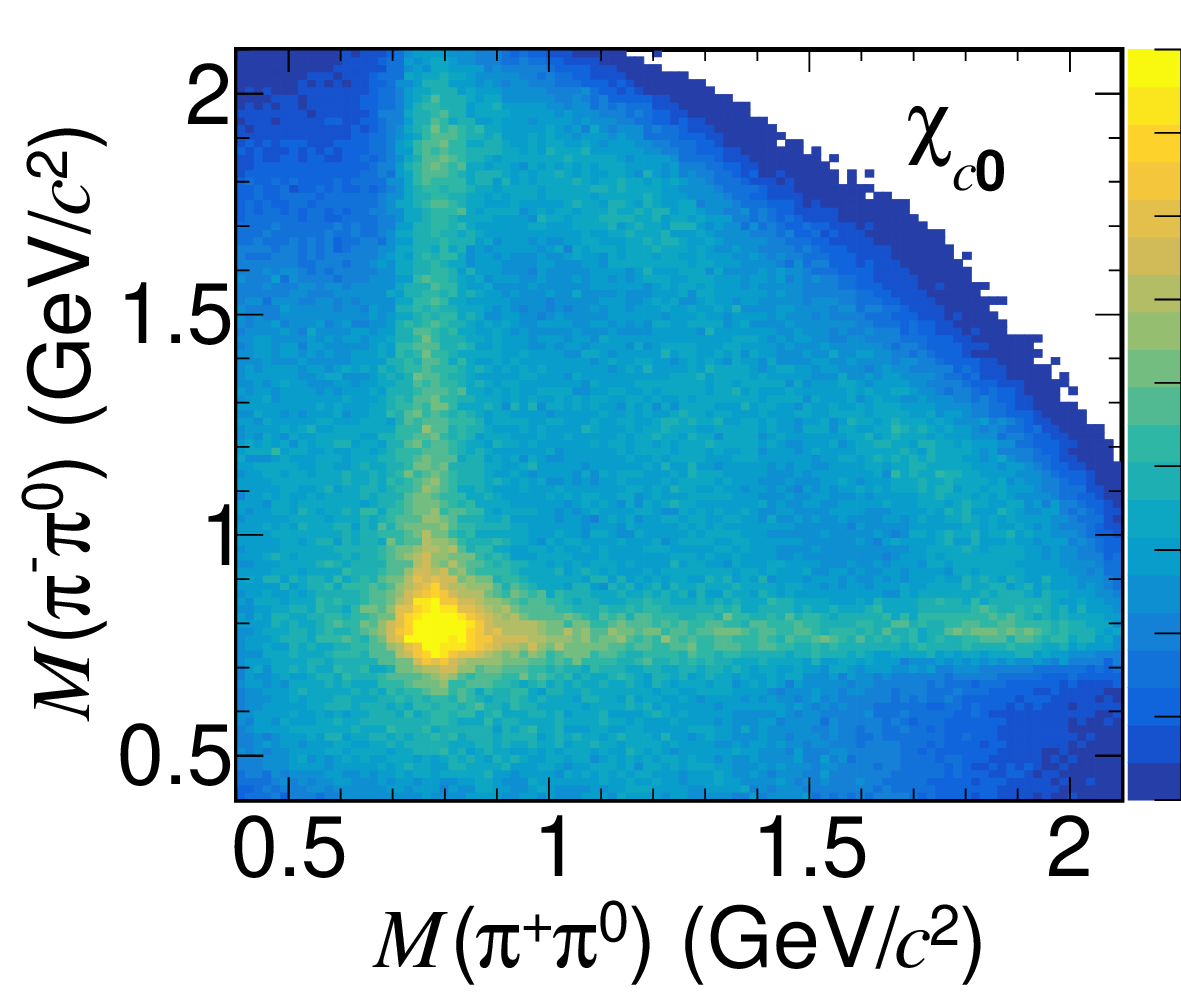}
\includegraphics[width=0.32\textwidth]{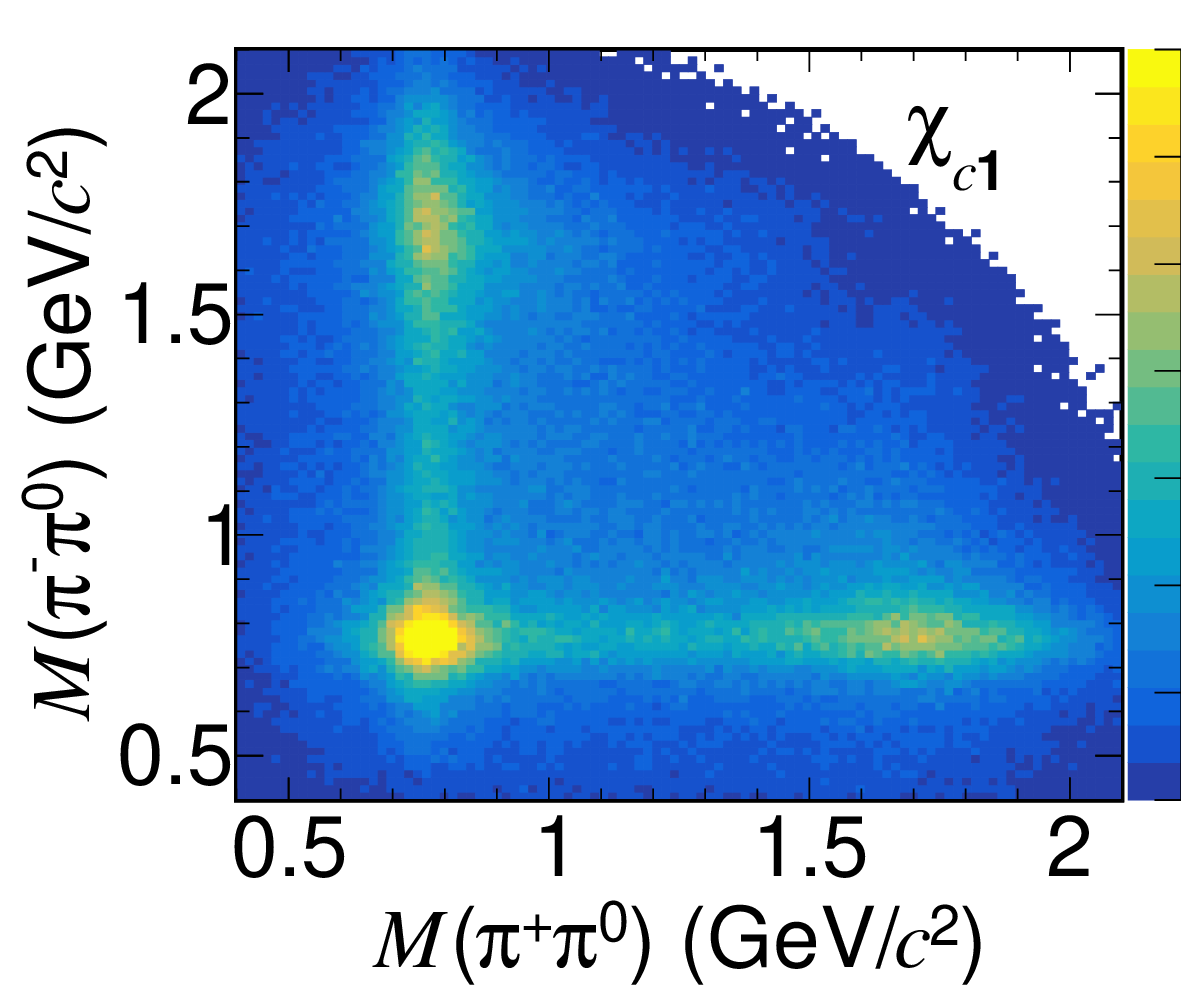}
\includegraphics[width=0.32\textwidth]{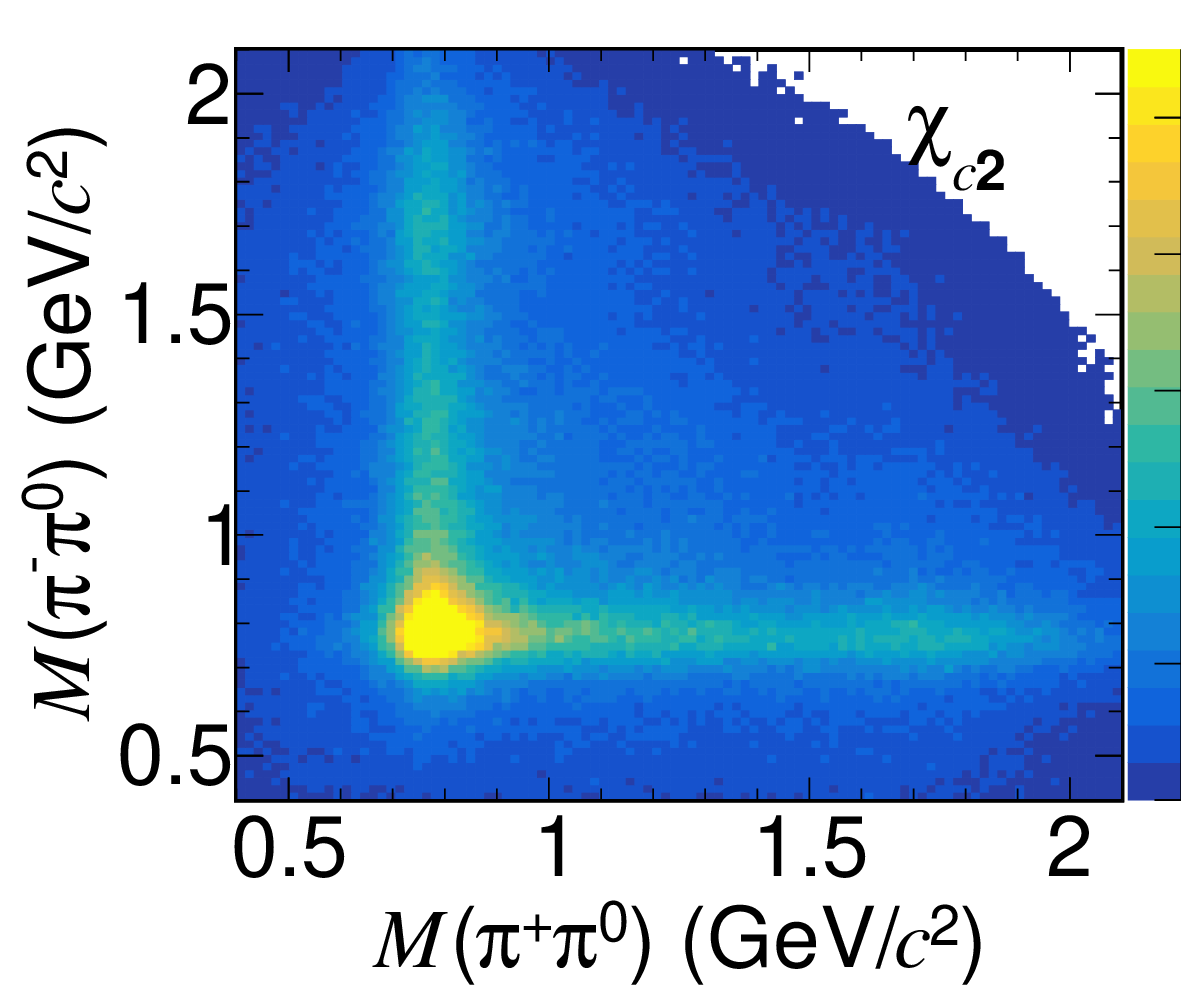}
\caption{The distribution of $M(\pi^+\pi^0)$ versus $M(\pi^-\pi^0)$ for events in the $\chi_{c0}$(left), $\chi_{c1}$(middle), and $\chi_{c2}$(right) signal regions.}
\label{1d2d}
\end{figure}

Figure~\ref{FIGDBA} shows the distribution of the invariant mass
$M(\pi^+\pi^-\pi^0\pi^0)$ with clear $\chi_{cJ}$
peaks. Figure~\ref{1d2d} shows the scatter plots of $M(\pi^+\pi^0)$
versus $M(\pi^-\pi^0)$ for events in the $\chi_{c0,1,2}$ mass
regions, where the \( \pi^+ \pi^0 \) and \( \pi^- \pi^0 \) combination
is chosen to minimize \( \Delta \equiv (M(\pi^+ \pi^0) - M_\rho)^2 +
(M(\pi^- \pi^0) - M_\rho)^2 \), with \( M_\rho \) being the nominal \(
\rho \) mass.  
The data indicates the presence of $\rho^+\rho^-$, 
$\rho\pi\pi$, and $\rho(1700)\pi\pi$ intermediate states.

\section {Signal yields and branching fraction calculation}

The $\chi_{cJ}$ yields are obtained by an unbinned maximum likelihood fit
to the $M(\pi^{+}\pi^{-}\pi^{0}\pi^{0})$ distribution. The probability
density function (PDF) employed in the fit includes the $\chi_{cJ}$ signals
 and five background functions.  The invariant mass
distributions from the exclusive MC samples are used to model the
$\chi_{cJ}$ signal shapes. Furthermore, to account for the
energy dependence in the $\mathrm{E1}$ transition vertex, the PDFs of
the signals are weighted event by event using the function
$\it{w}$~\cite{MCweight}, given by

\begin{equation}
w=\left(\frac{E_\gamma(M_t)}{E_{\text{on}}}\right)^3\times \mathrm{damp}(E_\gamma(M_t)),
\label{EQW}
\end{equation}
where $M_t$ denotes $M(\pi^{+}\pi^{-}\pi^{0}\pi^{0})$ from MC truth,
$E_{\gamma}(M_t) = \frac{m_{\psi(3686)}^{2}-M_t^{2}}{2m_{\psi(3686)}}$
is the energy of the transition photon in the rest frame of
$\psi$(3686), $m_{\psi(3686)}$ is the nominal mass of
$\psi$(3686), and $E_{\text{on}}$ is the fixed photon energy corresponding
to an on-shell $\chi_{cJ}$ decay, calculated as $E_{\text{on}} =
(m_{\psi(3686)}^{2} - m_{\chi_{cJ}}^{2})/(2m_{\psi(3686)})$ where
$m_{\chi_{cJ}}$ is the nominal $\chi_{cJ}$ mass from the PDG.
An empirical damping function adopted from the KEDR
experiment~\cite{KEDR} is used to control the infrared divergence in the
$E_\gamma^{3}$ factor and is given by
\begin{equation}
\mathrm{damp}(E_\gamma(M_t))=\frac{E_{\text{on}}^2}{E_\gamma E_{\text{on}}+\left(E_\gamma-E_{\text{on}}\right)^2}.
\label{EQDP1}
\end{equation} 

The signal PDF, $\text{PDF(s)}$,  is described by $\text{PDF}_w$, which is the
MC-simulated shape (not including the $w$ function) corrected for the
$w$-function effect, convolved with a Gaussian function ($G$) that accounts
for the resolution difference between data and MC simulation:
\begin{equation} \text{PDF(s)} = \text{PDF}_{w} \otimes G, \label
{PDFsignal} \end{equation} where $\otimes$ denotes convolution, and $G
\equiv G(m; \mu, \sigma)$ with $\mu$ and $\sigma$ as free
parameters in the fit.

The continuum background is estimated using data taken at
$\sqrt{s}=3.650$ GeV~\cite{xq365}.  To compensate for the phase-space
difference between $\sqrt{s}=3.650$ and $3.686$ GeV, the invariant mass $m$ is linearly scaled as
$m_{\text{shift}} = a(m - m_0) + m_0$, where $m_0$ is the production
threshold and $a$ adjusts for the energy shift from $\sqrt{s}=3.650$
GeV to $3.686$ GeV.  The line shape and number of events of the
peaking background component are fixed, as obtained from the inclusive
MC after normalization to the integrated luminosity of the
dataset. The background from $\psi(3686) \to \pi^0\pi^0 J/\psi$ is
modeled with its line shape fixed from the inclusive MC and the number
of events left as a free parameter in the fit.  The combinatorial
background comprises multiple processes and is not well described by a
single simulated shape. Therefore, we model it empirically with two
functions: a Gaussian function
approximates the relatively peaking component dominated by $\psi(3686)
\to \pi^0 h_{1}$ (with $h_{1} \to \rho\pi$), and an Argus
function~\cite{Argus} describes the remaining smooth,
threshold-dependent combinatorial background. The endpoint parameter
$m_0$ of the Argus function is fixed at 3.7 GeV, while its shape
parameter and overall normalization are determined from the fit to
data.

Figure~\ref{FIGDBA} shows the fit result, and the signal yields are
listed in Table~\ref{TABDSI}. The product branching fraction, defined
as $\mathcal{B}_{\mathrm{prod}} = \mathcal{B}(\psi(3686) \to
\gamma\chi_{cJ}) \cdot \mathcal{B}(\chi_{cJ} \to
\pi^{+}\pi^{-}\pi^{0}\pi^{0})$, is calculated as
\begin{equation}
\mathcal{B}_{\mathrm{prod}} = \dfrac{N^{\text{obs}}}{N^{\text{tot}}_{\psi(3686)} \cdot \mathcal{B}^{2}(\pi^{0} \to \gamma\gamma) \cdot \varepsilon_{\text{corr}}},
\label{EQDSI}
\end{equation}
where $N^{\text{obs}}$, $N^{\text{tot}}_{\psi(3686)}$, and
$\varepsilon_{\text{corr}}$ denote the number of net signal events,
the number of $\psi(3686)$ events, and the corrected detection
efficiency (corrected via the multidimensional reweighting defined in
Eq.~\ref{eq:efficiency_definition}), respectively.

\begin{table}[htbp]
\centering
\small
\begin{tabular}{l|ccc}
\hline
Source & $\chi_{c0}$ & $\chi_{c1}$ & $\chi_{c2}$ \\
\hline
$N^{\text{obs}}$ & $1363270 \pm 1436$ & $544611 \pm 876$ & $847320 \pm 1077$ \\
$\varepsilon_{\text{corr}}$ (\%) & $16.98 \pm 0.08$ & $18.19 \pm 0.08$ & $17.51 \pm 0.08$ \\
$\mathcal{B}_{\text{prod}}$ ($\times 10^{-3}$) & $3.03 \pm 0.01 \pm 0.12$ & $1.13 \pm 0.01 \pm 0.04$ & $1.83 \pm 0.01 \pm 0.07$ \\
$\mathcal{B}(\chi_{cJ} \to \pi^{+}\pi^{-}\pi^{0}\pi^{0})$ (\%) & $3.10 \pm 0.01 \pm 0.14$ & $1.16 \pm 0.01 \pm 0.05$ & $1.92 \pm 0.01 \pm 0.08$ \\
\hline
\end{tabular}
\caption{\label{TABDSI} Signal yields, detection efficiencies, and measured branching fractions.}
\end{table}

\section{Systematic uncertainty}

The sources of systematic uncertainties include the tracking and photon
efficiencies, $\pi^{0}$ reconstruction, PID, intermediate state,
kinematic fit, signal yields, background veto, number of
$\psi(3686)$ events, and quoted branching fractions.

\begin{itemize}
\item[$\bullet$] MDC Tracking and PID. The uncertainties in the MDC
tracking and PID for each pion are estimated using the control sample
$J/\psi \to\pi^{+}\pi^{-}\pi^{0}$, resulting in a systematic
uncertainty of $1.0\,\%$ for tracking and $1.0\,\%$ for PID per pion
track~\cite{sysMDC,sysMDC1}.

\item[$\bullet$] $\pi^{0}$ reconstruction. The uncertainty of the
$\pi^{0}$ reconstruction is studied with the control sample $J/\psi
\to \pi^{+}\pi^{-}\pi^{0}$, and determined to be 1.0\,$\%$ per
$\pi^{0}$ \cite{syspi0}, including the uncertainty from photon detection.

\item[$\bullet$] Photon detection. The uncertainty from photon
detection is 1\,$\%$, obtained from studies of the control sample
$J/\psi \to \gamma \pi^{0}\pi^{0}$ \cite{sysphoton}.

\item[$\bullet$] MC simulation. 
The presence of the intermediate states in 
Figure~\ref{1d2d} indicates that a pure PHSP model may not adequately capture the signal kinematics.
The uncertainty due to imperfections in the MC modeling of the $\cos\theta$ and momentum distributions of pion tracks (taking the $\chi_{c0}$  as an example in Figure~\ref{rewt0}) is estimated using a multi-variable product method, yielding systematic uncertainties of 0.2\,$\%$, 0.1\,$\%$, and 0.9\,$\%$ for $\chi_{c0}$, $\chi_{c1}$, and $\chi_{c2}$, respectively. 
The corrected efficiency $\varepsilon_{\text{corr}}$ is defined as:
\begin{equation}
 \begin{aligned}
 \varepsilon_{\text{corr}} &= \frac{\sum_{i,j,k} n_{i,j,k}^{\mathrm{MC}} \cdot \epsilon_{i,j,k}}{\sum_{i,j,k} n_{i,j,k}^{\mathrm{truth}} \cdot \epsilon_{i,j,k}}, \\[6pt]
 \epsilon_{i,j,k} &= 
 \begin{cases} 
 \dfrac{n_{i,j,k}^{\mathrm{data}} - n_{i,j,k}^{\mathrm{bkg}} \cdot f_{\mathrm{bkg}}}{n_{i,j,k}^{\mathrm{MC}}}, & n_{i,j,k}^{\mathrm{MC}} \neq 0 \\ 
 1, & n_{i,j,k}^{\mathrm{MC}} = 0 
 \end{cases}.
 \end{aligned}
 \label{eq:efficiency_definition}
\end{equation}
where the indices $(i, j, k)$ denote the bins of $\cos\theta$, momentum, and particle type, respectively. $n^{\mathrm{truth}}$ and $n^{\mathrm{MC}}$ are the event yields from the MC generator and after full simulation. $n^{\mathrm{data}}$ and $n^{\mathrm{bkg}}$ are yields from data and estimated background in the control sample. $f_{\mathrm{bkg}}$ is the background fraction. The factor $\epsilon_{i,j,k}$ is the bin-wise data/MC efficiency ratio.

\begin{figure}[htbp]
\centering
\includegraphics[width=0.97\textwidth]{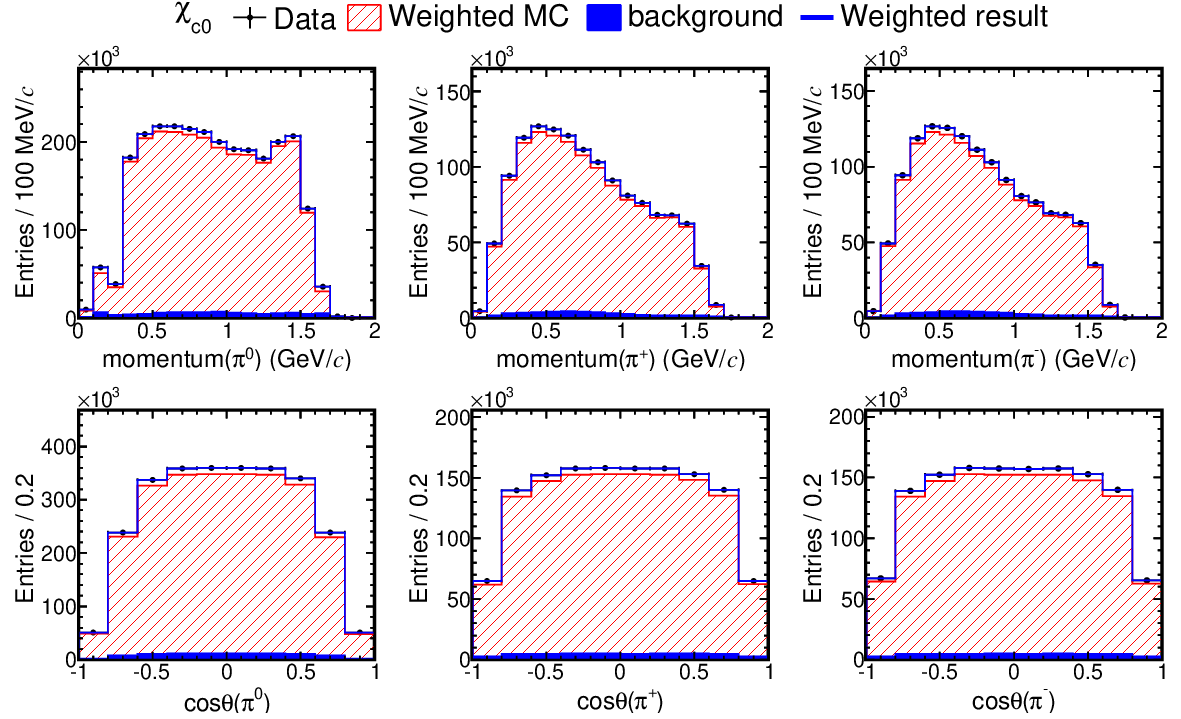}
\caption{Comparison of momentum and $\cos\theta$ distributions for $\pi^0$, $\pi^+$, and $\pi^-$ between data and signal MC samples.
From top to bottom, the rows show the distributions for the momentum and $\cos\theta$ distributions, respectively. From left to right, the columns show  the $\pi^0$, $\pi^+$, and $\pi^-$, respectively.
The $\chi_{c0}$ signal MC is corrected to match the data using the multivariable product method. The black points with error bars represent the data, and the red lines show the corrected MC distributions. The $\chi_{c0}$ distributions are shown, as an example.}
\label{rewt0}
\end{figure}

\item[$\bullet$] Intermediate states. 
A study based on a two-dimensional fit to data, which constrains the
possible double-$\rho$ contribution (modeled with the Gounaris-Sakurai
line shape~\cite{GSmodle}), shows that the resulting efficiency
variations (0.1\,$\%$, 0.3\,$\%$, 0.4\,$\%$ for $\chi_{cJ}$) are of
the same order as the MC uncertainties from pion reweighting.
Therefore, no additional uncertainty is assigned to avoid double-counting.

\item[$\bullet$] Kinematic fit. The systematic uncertainty due to the kinematic fit 
is estimated by comparing the detection 
efficiencies with and without correcting track helix parameter~\cite{KinematicFit1}.

\item[$\bullet$] $J/\psi$- or $K^{0}_{S}K^{0}_{S}$-related background
veto. The uncertainties caused by these two background vetoes are
estimated by varying their mass windows in steps of 5 MeV/$c^2$. For
each step, the deviation between the alternative and nominal branching
fraction is evaluated using the Barlow
test~\cite{barlow,barlow1,barlow2} statistic $\zeta = |B_{\text{nom}}
- B_{\text{test}}| / \sqrt{|\sigma_{\text{nom}}^2 -
\sigma_{\text{test}}^2|}$, where $B$ denotes the branching fraction
and $\sigma$ is the statistical uncertainty.  All $\zeta$ values obtained
are found to be less than 2.0, indicating negligible systematic
effects according to the Barlow criterion. Therefore, no 
uncertainty is assigned for these two sources.

\item[$\bullet$] Signal yields. 
The systematic uncertainty in signal yields, obtained by fitting the $M(\pi^+\pi^-\pi^0\pi^0)$ spectrum,  includes contributions from the background function, the peaking background, the fit range and the signal function. They are estimated as follows:
    \begin{itemize}
        \item[$-$] The uncertainty caused by the background function is estimated by comparing the fits where the background is modeled by an Argus function only and by an Argus function plus a 2nd order Chebyshev polynomial.
        \item[$-$] The uncertainty from peaking backgrounds is estimated by performing two types of fits with its yield fixed and floated, respectively. The resulting difference in the signal yield is taken as the systematic uncertainty.
        \item[$-$] The uncertainty from the fit range is estimated by performing multiple variations of the range boundaries. Both the lower and upper limits are independently adjusted in steps of 0.5\,MeV/$c^2$, with four shifts in each direction within $\pm$\,2.5~MeV/$c^2$ around the nominal range of [3.25, 3.65]~GeV/$c^2$. For each variation, the Barlow test statistic $\zeta$ is calculated, and all values obtained are found to be below 2.0, indicating a negligible systematic uncertainty.

        \item[$-$] The uncertainty caused by the damping function is
        estimated by replacing it with $e^{-E_\gamma^2 / (8\beta^2)}$,
        which was used by CLEO~\cite{CLEO} 
        to describe the dynamics of the production vertex from
        $\psi(3686)$. The differences on the $\chi_{cJ}$ signal yields
        are taken as the systematic uncertainties. This results in
        relative systematic uncertainties of 1.1\,\%, 0.2\,\%, and
        0.1\,\% for the $\chi_{c0}$, $\chi_{c1}$, and $\chi_{c2}$
        branching fractions, respectively, which are the dominant
        contributions to the total uncertainties of the signal yields.
    \end{itemize}

\item[$\bullet$] Number of $\psi(3686)$. The uncertainty in the number of $\psi(3686)$ events is determined to be 0.5\,$\%$ with inclusive hadronic $\psi(3686)$ decays~\cite{allpsi}.

\item[$\bullet$] MC statistics. The systematic uncertainties due to the limited statistics of the signal MC samples are 0.5\,$\%$, 0.4\,$\%$, and 0.4\,$\%$ for $\chi_{cJ} \to \pi^{+}\pi^{-}\pi^{0}\pi^{0}$, respectively. They are calculated via $\frac{\sqrt{(1 - \varepsilon) / N}}{\varepsilon}$, where $\varepsilon$ is the detection efficiency and $N$ is the total number of MC events.

\item[$\bullet$] Branching fractions of $\psi(3686) \to \gamma\chi_{cJ}$. The branching fractions for $\chi_{c0}$, $\chi_{c1}$, and $\chi_{c2}$ are taken from the PDG~\cite{PDG} and their uncertainties are 2.0\,\%, 2.5\,\%, and 2.1\,\%, respectively.
\end{itemize}

Table \ref{TABDSHISI} lists all of the systematic uncertainties. 
The total systematic uncertainities are obtained from quadrature sum of individual systematic uncertainties.  

\begin{table}
\centering
\normalsize
\setlength{\tabcolsep}{12pt} 
\begin{tabular}{l c c c}
\hline
Source&$\chi_{c0}$&$\chi_{c1}$&$\chi_{c2}$\\
\hline
Tracking&2.0&2.0&2.0\\
PID&2.0&2.0&2.0\\
$\pi^{0}$ reconstruction&2.0&2.0&2.0\\
Photon detection&1.0&1.0&1.0\\
MC simulation&0.2&0.1&0.9\\
Intermediate state&n&n&n\\
Kinematic fit&0.3&0.6&0.6\\
$J/\psi$ veto&n&n&n\\
$K^0_{\mathrm{S}}K^0_{\mathrm{S}}$ veto&n&n&n\\
Signal yields&1.1&0.3&0.1\\
$N_{\psi(3686)}$&0.5&0.5&0.5\\
MC statistics&0.5&0.4&0.4\\
\hline
$\mathcal{B}$($\psi(3686)\to\gamma\chi_{cJ}$)$\cdot$$\mathcal{B}$($\chi_{cJ}\to\pi^{+}\pi^{-}\pi^{0}\pi^{0}$)&3.9&3.7&3.8\\
\hline
$\mathcal{B}$($\chi_{cJ}\to\pi^{+}\pi^{-}\pi^{0}\pi^{0}$)&4.4&4.5&4.4\\
\hline
\end{tabular}
\caption{\label{TABDSHISI} Relative systematic uncertainties ($\%$) for $\chi_{cJ} \to \pi^{+}\pi^{-}\pi^{0}\pi^{0}$, where "n" means negligible.}
\end{table}

\section{Summary}

Using $(2712.4\,\pm\,14.3)\times10^{6}$ $\psi(3686)$ events, we report
significantly improved measurements of the $\chi_{cJ} \to
\pi^+\pi^-\pi^0\pi^0$ branching fractions through $\psi(3686) \to
\gamma\chi_{cJ}$ decays, superseding previous results by BESIII. The measured branching fractions are: $\mathcal{B}(\chi_{c0}
\to \pi^{+}\pi^{-}\pi^{0}\pi^{0}) = (3.10 \pm 0.01 \pm 0.14) \times
10^{-2}$; $\mathcal{B}(\chi_{c1} \to \pi^{+}\pi^{-}\pi^{0}\pi^{0}) =
(1.16 \pm 0.01 \pm 0.05) \times 10^{-2}$; and $\mathcal{B}(\chi_{c2}
\to \pi^{+}\pi^{-}\pi^{0}\pi^{0}) = (1.92 \pm 0.01 \pm 0.08) \times
10^{-2}$, where the first uncertainties are statistical and the second
systematic.  Our measurements can be directly compared with the
previous most precise results from the CLEO
collaboration~\cite{2008CLEO}. Our results are in good agreement
with those values but are determined with a statistical precision
improved by nearly an order of magnitude.

The neutral-to-charged ratios $R_J=\mathcal{B}(\chi_{cJ}\to\pi^+\pi^-\pi^0\pi^0)/\mathcal{B}(\chi_{cJ}\to2\pi^+2\pi^-)$\ are calculated to be $1.57 \pm 0.11$ for $J = 0$, $1.70 \pm 0.11$ for $J = 1$, and $1.70 \pm 0.12$ for $J = 2$. 
Although these ratios provide significant constraints, their interpretation is challenging due to the complex dynamics of multi-pion final states, which involve both QCD and QED processes as well as non-trivial angular-momentum couplings. 
The analysis, enabled by the improved precision, indicates that $\rho^+\rho^-$ intermediate states are dominant in the $\chi_{cJ}\to\pi^+\pi^-\pi^0\pi^0$ decays.

\acknowledgments
\input{acknowledgement_2025-12-05}

\appendix
\section{authorlist}
\input{authorlist_2025-12-05}
\end{document}

%% file: acknowledgement_2025-12-05.tex

The BESIII Collaboration thanks the staff of BEPCII (https://cstr.cn/31109.02.BEPC) and the IHEP computing center for their strong support. This work is supported in part by National Key R\&D Program of China under Contracts Nos. 2025YFA1613900, 2023YFA1606000, 2023YFA1606704; National Natural Science Foundation of China (NSFC) under Contracts Nos. 11635010, 11935015, 11935016, 11935018, 12025502, 12035009, 12035\\013, 12061131003, 12192260, 12192261, 12192262, 12192263, 12192264, 12192265, 12221005, 12225509, 12235017, 12342502, 12361141819; the Chinese Academy of Sciences (CAS) Large-Scale Scientific Facility Program; the Strategic Priority Research Program of Chinese Academy of Sciences under Contract No. XDA0480600; CAS under Contract No. YSBR-101; 100 Talents Program of CAS; The Institute of Nuclear and Particle Physics (INPAC) and Shanghai Key Laboratory for Particle Physics and Cosmology; ERC under Contract No. 758462; German Research Foundation DFG under Contract No. FOR5327; Istituto Nazionale di Fisica Nucleare, Italy; Knut and Alice Wallenberg Foundation under Contracts Nos. 2021.0174, 2021.0299, 2023.0315; Ministry of Development of Turkey under Contract No. DPT2006K-120470; National Research Foundation of Korea under Contract No. NRF-2022R1A2C1092335; National Science and Technology fund of Mongolia; Polish National Science Centre under Contract No. 2024/53/B/ST2/00975; STFC (United Kingdom); Swedish Research Council under Contract No. 2019.04595; U. S. Department of Energy under Contract No. DE-FG02-05ER41374.


%% file: authorlist_2025-12-05.tex
M.~Ablikim$^{1}$\BESIIIorcid{0000-0002-3935-619X},
M.~N.~Achasov$^{4,d}$\BESIIIorcid{0000-0002-9400-8622},
P.~Adlarson$^{82}$\BESIIIorcid{0000-0001-6280-3851},
X.~C.~Ai$^{88}$\BESIIIorcid{0000-0003-3856-2415},
C.~S.~Akondi$^{31A,31B}$\BESIIIorcid{0000-0001-6303-5217},
R.~Aliberti$^{39}$\BESIIIorcid{0000-0003-3500-4012},
A.~Amoroso$^{81A,81C}$\BESIIIorcid{0000-0002-3095-8610},
Q.~An$^{78,64,\dagger}$,
Y.~H.~An$^{88}$\BESIIIorcid{0009-0008-3419-0849},
Y.~Bai$^{62}$\BESIIIorcid{0000-0001-6593-5665}
% ...
\\
\vspace{0.2cm}
(BESIII Collaboration)\\
\vspace{0.2cm} {\it
$^{1}$ Institute of High Energy Physics, Beijing 100049, People's Republic of China\\
% ...
$^{4}$ Budker Institute of Nuclear Physics SB RAS (BINP), Novosibirsk 630090, Russia\\
% ...
$^{31}$ INFN Sezione di Ferrara, (A)INFN Sezione di Ferrara, I-44122, Ferrara, Italy; (B)University of Ferrara, I-44122, Ferrara, Italy\\
% ...
$^{39}$ Johannes Gutenberg University of Mainz, Johann-Joachim-Becher-Weg 45, D-55099 Mainz, Germany\\
% ...
$^{62}$ Southeast University, Nanjing 211100, People's Republic of China\\
% ...
$^{64}$ State Key Laboratory of Particle Detection and Electronics, Beijing 100049, Hefei 230026, People's Republic of China\\
% ...
$^{78}$ University of Science and Technology of China, Hefei 230026, People's Republic of China\\
% ...
$^{81}$ University of Turin and INFN, (A)University of Turin, I-10125, Turin, Italy; (B)University of Eastern Piedmont, I-15121, Alessandria, Italy; (C)INFN, I-10125, Turin, Italy\\
$^{82}$ Uppsala University, Box 516, SE-75120 Uppsala, Sweden\\
% ...
$^{88}$ Zhengzhou University, Zhengzhou 450001, People's Republic of China\\
% ...
$^{\dagger}$ Deceased\\
% ...
$^{d}$ Also at the Novosibirsk State University, Novosibirsk, 630090, Russia\\
% ...

}

%% file: bam954_FH_CL.bbl
\begin{thebibliography}{99}

\bibitem{bes3review2022}
W.~Song,
\textit{Study of Charmonium(-like) Spectroscopy and Decay at BESIII},
\href{https://doi.org/10.48550/arXiv.2209.15061}{arXiv:2209.15061 [hep-ex]}.

\bibitem{2008CLEO}
CLEO Collaboration,
\textit{First observation of exclusive $\chi_{cJ}$ decays to two charged and two neutral hadrons},
\href{https://doi.org/10.1103/PhysRevD.78.092004}{Phys.\ Rev.\ D \textbf{78} (2008) 092004}.

\bibitem{1997HWHuang}
H.~W.~Huang and K.~T.~Chao, 
\textit{QCD predictions for annihilation decays of $P$-wave quarkonia to next-to-leading order in $\alpha_s$}, 
\href{https://doi.org/10.1103/PhysRevD.56.1821}{Phys.\ Rev.\ D \textbf{56} (1997) 1821}.

\bibitem{1996APetrelli}
A.~Petrelli,
\textit{Colour-octet NLO QCD corrections to hadronic $\chi_{cJ}$ decays},
\href{https://doi.org/10.1016/0370-2693(96)00459-5}{Phys.\ Lett.\ B \textbf{380} (1996) 159}.

\bibitem{1997JBolz}
J.~Bolz, P.~Kroll and G.~A.~Schuler,
\textit{Colour-octet contributions to exclusive charmonium decays},
\href{https://doi.org/10.1016/S0370-2693(96)01515-8}{Phys.\ Lett.\ B \textbf{392} (1997) 198}.

\bibitem{2000SMHWong1}
S.~M.~H.~Wong,
\textit{Color octet contribution in exclusive $P$-wave charmonium decay into octet and decuplet baryons},
\href{https://link.springer.com/article/10.1007/s100520000376}{Eur.\ Phys.\ J.\ C \textbf{14} (2000) 643}.

\bibitem{PDG}
Particle Data Group,
\textit{Review of Particle Physics},
\href{https://doi.org/10.1103/PhysRevD.110.030001}{Phys. Rev. D \textbf{110} (2024) 030001}.

\bibitem{allpsi}
BESIII Collaboration,
\textit{Determination of the number of $\psi(3686)$ events at BESIII},
\href{https://doi.org/10.1088/1674-1137/ad595b}{Chin.\ Phys.\ C \textbf{48} (2024) 093001}.

\bibitem{besiii}
BESIII~Collaboration,
\textit{Design and construction of the BESIII detector},
\href{https://doi.org/10.1016/j.nima.2009.12.050}{Nucl.\ Instrum.\ Meth.\ Phys.\ Res.\ Sect.\ A \textbf{614} (2010) 345}.

\bibitem{bepcii} 
C.~H.~Yu \textit{et al.},
\textit{BEPCII Performance and Beam Dynamics Studies on Luminosity},
in \textit{Proc. 7th Int. Particle Accelerator Conf.} (IPAC'16),
\href{https://accelconf.web.cern.ch/ipac2016/doi/JACoW-IPAC2016-TUYA01.html}{Busan, Korea, May 2016,  
pp.~1342--1345}.

\bibitem{MDC}
J.~K.~Wang \textit{et al.},
\textit{BESIII track fitting algorithm},
\href{https://doi.org/10.1088/1674-1137/33/10/010}
{Chin.\ Phys.\ C \textbf{33} (2009) 870--879}.

\bibitem{etof1}
X.~Li \textit{et al.},
\textit{Study of MRPC technology for BESIII endcap-TOF upgrade},
\href{https://doi.org/10.1007/s41605-017-0014-2}{Radiat.\ Detect.\ Technol.\ Methods \textbf{1} (2017) 13}.
 
\bibitem{etof2}
Y.~X.~Guo \textit{et al.},
\textit{The study of time calibration for upgraded end cap TOF of BESIII},
\href{https://doi.org/10.1007/s41605-017-0012-4}{Radiat.\ Detect.\ Technol.\ Methods \textbf{1} (2017) 15}.
 
\bibitem{etof3} 
P.~Cao \textit{et al.},
\textit{Design and construction of the new BESIII endcap Time-of-Flight system with MRPC Technology},
\href{https://doi.org/10.1016/j.nima.2019.163053}{Nucl.\ Instrum.\ Meth.\ A \textbf{953} (2020) 163053}.

\bibitem{boos}
W.~D.~Li \textit{et al.},
\textit{The BESIII Offline Software},
in \textit{Proc. Computing in High Energy and Nuclear Physics Conf.} (CHEP'06), 
\href{https://indico.cern.ch/event/408139/contributions/979815/}{Mumbai, India, February 2006}

\bibitem{geant4}
GEANT4 Collaboration,
\textit{GEANT4-A simulation toolkit},
\href{https://doi.org/10.1016/S0168-9002(03)01368-8}{Nucl.\ Instrum.\ Meth.\ A \textbf{506} (2003) 250--303}.

\bibitem{Boost}
Z.~Y.~Deng, G.~F.~Cao \textit{et al.},
\textit{Object-Oriented BESIII Detector Simulation System},
\href{http://hepnp.ihep.ac.cn/article/id/283d17c0-e8fa-4ad7-bfe3-92095466def1}{Chin.\ Phys.\ C \textbf{30} (2006) 371--376}.

\bibitem{gtheta01}
W.~M.~Tancnbaum $et~al.,$
\textit{Radiative decays of the $\psi(3684)$ into high-mass states},
\href{https://doi.org/10.1103/PhysRevD.17.1731}{Phys. Rev. D \textbf{17} (1978) 1731}.

\bibitem{BesEvt}
D.~J.~Lange,
\textit{The EvtGen particle decay simulation package},
\href{https://doi.org/10.1016/S0168-9002(01)00089-4}{Nucl.\ Instrum.\ Meth.\ A \textbf{462} (2001) 152--155}.

\bibitem{KKMC1}
S.~Jadach, B.~F.~L.~Ward and Z.~Was,
\textit{The precision Monte Carlo event generator KK for two-fermion final states in $e^+e^-$ collisions},
\href{https://doi.org/10.1016/S0010-4655(00)00048-5}{Comp.\ Phys.\ Commun.\ \textbf{130} (2000) 260--311}.

\bibitem{KKMC2}
S.~Jadach, B.~F.~L.~Ward and Z.~Was,
\textit{Coherent exclusive exponentiation for precision Monte Carlo calculations},
\href{https://doi.org/10.1103/PhysRevD.63.113009}{Phys.\ Rev.\ D \textbf{63} (2001) 113009}.

\bibitem{LUNDCHARM}
R.~G.~Ping,
\textit{Event generators at BESIII},
\href{https://doi.org/10.1088/1674-1137/32/8/001}{Chin.\ Phys.\ C \textbf{32} (2008) 599--604}.

\bibitem{MCweight}
BESIII Collaboration,
\textit{Search for  $ \eta_c(2S) \to p\bar{p}K^+K^- $  and measurement of  $ \chi_{cJ} \to p\bar{p}K^+K^- $  in  $ \psi(3686) $  radiative decays},
\href{https://doi.org/10.1103/PhysRevD.111.072001}{Phys. Rev. D \textbf{111} (2025) 072001}.

\bibitem{KEDR}
KEDR Collaboration,
\textit{Measurement of $B(J/\psi \to \eta_c\gamma)$ with the KEDR},
\href{https://doi.org/10.1142/S2010194511000791}{Int.\ J.\ Mod.\ Phys.\ Conf.\ Ser.\ \textbf{02} (2011) 188--192.}

\bibitem{xq365}
BESIII Collaboration,
\textit{ Observation of $\eta_{c}(1S,2S)$ and $\chi_{cJ}$ decays
 to $2(\pi^+\pi^-)\eta$, via $\psi(3686)$ radiative transitions},
\href{https://doi.org/10.1103/PhysRevD.111.052013}{Phys. Rev. D \textbf{111} (2025) 052013}.

\bibitem{Argus}
ARGUS Collaboration,
\textit{Search for hadronic $b \to u$ decays},
\href{https://doi.org/10.1016/0370-2693(90)91293-K}{Phys.\ Lett.\ B \textbf{241} (1990) 278--282.}

\bibitem{sysMDC}
BESIII Collaboration,
\textit{Study of $\chi_{cJ}$ radiative decays into a vector meson},
\href{https://doi.org/10.1103/PhysRevD.83.112005}{Phys.\ Rev.\ D \textbf{83} (2011) 112005}.

\bibitem{sysMDC1}
BESIII Collaboration,
\textit{Observation of $\eta' \to \pi^{+}\pi^{-}\pi^{+}\pi^{-}$ and $\eta' \to \pi^{+}\pi^{-}\pi^{0}\pi^{0}$},
\href{https://doi.org/10.1103/PhysRevLett.112.251801}{Phys.\ Rev.\ L \textbf{113} (2014) 039903}.

\bibitem{syspi0}
BESIII Collaboration,
\textit{Branching fraction measurements of $\chi_{c0}$ and $\chi_{c1}$ to $\pi^0\pi^0$ and $\eta\eta$},
\href{https://doi.org/10.1103/PhysRevD.81.052005}{Phys.\ Rev.\ D \textbf{81} (2010) 052005}.

\bibitem{sysphoton}
BESIII Collaboration,
\textit{Amplitude analysis of the $\pi^0\pi^0$ system produced in radiative $J/\psi$ decays},
\href{https://doi.org/10.1103/PhysRevD.92.052003}{Phys.\ Rev.\ D \textbf{92} (2016) 052003}.

\bibitem{GSmodle}
G.~J.~Gounaris and J.~J.~Sakurai,
\textit{Finite-width corrections to the vector-meson-dominance prediction for $\rho\to e^+e^-$},
\href{https://doi.org/10.1103/PhysRevLett.21.244}{Phys.\ Rev.\ Lett.\ \textbf{21} (1968) 244--246}.

\bibitem{KinematicFit1}
BESIII Collaboration,
\textit{Search for hadronic transition and observation of $\chi_{cJ} \to KK\pi\pi\pi$},
\href{https://doi.org/10.1103/PhysRevD.87.012002}{Phys.\ Rev.\ D \textbf{87} (2013) 012002}.

\bibitem{barlow}
R.~J.~Barlow,
\textit{Systematic Errors: facts and fictions},
\href{https://doi.org/10.48550/arXiv.hep-ex/0207026}{arXiv:hep-ex/0207026}.

\bibitem{barlow1}
O.~Behnke, K.~Kr\"oninger, G.~Schott and T.~Sch\"orner-Sadenius,
\textit{Data Analysis in High Energy Physics: A Practical Guide to Statistical Methods},
(Wiley-VCH, Berlin, Germany, 2013),
\href{https://doi.org/10.1002/9783527653416}{10.1002/9783527653416}.

\bibitem{barlow2}
BESIII Collaboration,
\textit{Probing CP symmetry and weak phases with entangled double-strange baryons},
\href{https://doi.org/10.1038/s41586-022-04624-1}{Nature \textbf{606} (2022) 64–68}.

\bibitem{CLEO}
CLEO Collaboration,
\textit{Precision Measurement of the $\eta_c(1S)$ Mass and Width},
\href{https://doi.org/10.1103/PhysRevLett.102.011801}{Phys.\ Rev.\ Lett.\ \textbf{102} (2009) 011801.} 
\end{thebibliography}
